\pdfoutput=1
\documentclass[11pt]{article}
\usepackage{graphicx}
\usepackage{amssymb}
\usepackage{graphics}

\input{tcilatex}
\begin{document}

\title{Constraining spin-dependent short range interactions using neutron
spin precession close to a mirror}
\author{O. Zimmer \\
Institut Laue Langevin, 38042 Grenoble, France, \\
and Physik-Department E18, TU M\"{u}nchen, 85748 Garching, Germany}
\maketitle

\begin{abstract}
Spin-dependent short range interactions of free neutrons with matter may be
searched for in various ways. This short note discusses pseudomagnetic
precession of trapped ultracold neutrons in vicinity to bulk matter, which
should be several orders of magnitude more sensitive than any other method
proposed so far.

\medskip

\bigskip
\end{abstract}

Spin-dependent short-range interactions may be induced by light,
pseudoscalar bosons such as the axion invented to solve the strong CP
problem \cite{Moody/1984}. These hypothetic particles are proposed to
mediate a parity and time reversal violating interaction between a fermion
and the spin of another fermion, which is parameterised by a Yukawa-type
potential with range $\lambda $ and with a monopole-dipole coupling.
Considering a neutron with mass $m_{\mathrm{n}}$ and spin $\frac{1}{2}%
\hslash \mathbf{\sigma }$ interacting with another nucleon at distance $%
\mathbf{r}$ it may be written as 
\begin{equation}
V\left( \mathbf{r}\right) =\kappa \mathbf{n\cdot \sigma }\left( \frac{1}{%
\lambda r}+\frac{1}{r^{2}}\right) \mathrm{e}^{-r/\lambda },\qquad \kappa =%
\frac{\hslash ^{2}g_{\mathrm{S}}g_{\mathrm{P}}}{8\pi m_{\mathrm{n}}},
\label{V-sp}
\end{equation}%
with unitless scalar and pseudo-scalar coupling constants $g_{\mathrm{S}}$
and $g_{\mathrm{P}}$ between the neutron and the exchanged boson. $\mathbf{n}%
=\mathbf{r}/r$ is a unit distance vector from the neutron to the nucleon.
Prior experiments and astronomical observations suggest that, if the axion
exists, its mass should lie within the "axion window" $1$ $\mathrm{\mu eV}%
\lesssim m_{\mathrm{A}}\lesssim 1$ meV, corresponding to a range $0.2$ $%
\mathrm{mm}\lesssim \lambda \lesssim 20$ mm.

A first limit on such an interaction was provided in \cite{Baessler/2007}
along with a proposal for further improvement using gravitationally bound
quantum states of the free neutron. The suggested Stern-Gerlach type
experiment employs neutron transmission through a slit between an absorber
and a horizontal mirror, which is sensitive to the shape of the spatial
neutron wave function. The hypothetic short-range forces of mirror and
absorber would induce a spin dependence. In this short note a spin
precession experiment is proposed that should be more sensitive by several
orders of magnitude.

Consider a neutron close to the surface of a massive plane mirror with
thickness $d$. Let $z$ denote the coordinate normal to the mirror with $z=0$
at the surface. The effective potential of the mirror with nucleon number
density $N$ is obtained by integration of eq.(\ref{V-sp}) over the mirror
volume and is given by\footnote{%
Equation (\ref{V-eff}) holds also inside the mirror. The potential is
sizeable only for $\left\vert z\right\vert \lesssim \lambda $, in contrast
to spin-independent short-range interactions, for which the potential
attains its maximum value inside the mirror \cite{Zimmer/2006}.}%
\begin{equation}
V\left( z\right) =V\left( 0\right) \left( \mathrm{e}^{-\left\vert
z\right\vert /\lambda }-\mathrm{e}^{-\left\vert z+d\right\vert /\lambda
}\right) \sigma _{z},\qquad V\left( 0\right) =2\pi N\kappa \lambda .
\label{V-eff}
\end{equation}%
With respect to spin dependence it has the same analytic form as the
interaction of the neutron magnetic moment with a magnetic field pointing in 
$z$ direction. $V\left( z\right) $ can therefore be probed by searching for
a pseudomagnetic precession of neutrons polarised parallel to the surface of
the mirror. For this investigation Ramsey's resonance method applied to
ultracold neutrons (UCN) is particularly well suited. The proposed
experiment requires only slight modification of devices employed in ongoing
searches for the electric dipole moment (EDM) of the neutron which have
matured to sensitivity beyond $10^{-21}$ eV.

Here we consider spin precession of UCN stored in a trap made of two plane
mirrors with distance $D$ and with a large difference in mass density. Let
the mirror surfaces be located at $z=0$ and $z=D$, and a homogeneous
magnetic field $\mathbf{B}$ be applied in $z$ direction. For maximum
strength of the potential $V\left( z\right) $ the thickness of the mirror
with high mass density is to be chosen $d\gg \lambda $. The trapped neutrons
sense the spatially averaged spin-dependent potential, which, neglecting the
influence of the quantum-mechanical boundary conditions on the probability
density close to the mirror, and also a small contribution due to the light
mirror, is given by%
\begin{equation}
\overline{V}=\pm \frac{1}{D}\int_{0}^{D}V\left( z\right) \mathrm{d}z=\pm
V\left( 0\right) \frac{\lambda }{D}\left( 1-\mathrm{e}^{-D/\lambda }\right)
\left( 1-\mathrm{e}^{-d/\lambda }\right) ,  \label{V-average}
\end{equation}%
with the signs for spin parallel and anti-parallel to $\mathbf{B}$. Due to
the operator $\mathbf{n\cdot \sigma }$ in eq.(\ref{V-sp}) the signs get
inverted under inversion of the trap orientation relative to $\mathbf{B}$.
Hence, neglecting for the moment the issue of magnetic field instability, we
may determine $\overline{V}$ from the difference%
\begin{equation}
\omega _{+}-\omega _{-}=2\left\vert \overline{V}\right\vert /\hslash
\label{effect}
\end{equation}%
of neutron precession frequencies for the two trap orientations with respect
to the magnetic field.

The counting statistical uncertainty of this basic measuring procedure is
given by (see, e.g. ref. \cite{Pendlebury/2000} for the analogue EDM search)%
\begin{equation}
\sigma \left( \overline{V}\right) =\frac{\hslash }{2\alpha T\sqrt{N_{\mathrm{%
n}}}},
\end{equation}%
where $N_{\mathrm{n}}$ is the total number of UCNs counted in a series of
measurement cycles, $T$ is the time of UCN storage per cycle, and $\alpha $
denotes the visibility of the Ramsey fringes. The new interaction will thus
be detectable if%
\begin{equation}
g_{\mathrm{S}}g_{\mathrm{P}}\geq \frac{2m_{\mathrm{n}}}{\alpha TN\hslash 
\sqrt{N_{\mathrm{n}}}}\frac{D}{\lambda ^{2}}\left( 1-\mathrm{e}^{-D/\lambda
}\right) ^{-1}\left( 1-\mathrm{e}^{-d/\lambda }\right) ^{-1}.
\end{equation}%
For a conservative estimate of the statistical sensitivity we consider
employing one of the existing devices to search for the neutron EDM, equiped
with a mirror made from lead\footnote{%
Although lead is not too bad for neutron storage, a thin coating with a
material with large Fermi potential and low UCN loss per wall collision will
improve $T$ and $N_{\mathrm{n}}$. Its lower mass density reduces sensitivity
only for small $\lambda $ in the $\mathrm{\mu m}$ range out of the axion
window.} ($N=6.86\times 10^{30}$ m$^{-3}$), and taking $T=100$ s, and $N_{%
\mathrm{n}}=10^{8}$ (attainable during one reactor cycle at the present UCN
source at the ILL). To allow for imperfect polarisation efficiency we set $%
\alpha =0.5$. With $D=0.1$ m we thus might detect a signal if%
\begin{equation}
g_{\mathrm{S}}g_{\mathrm{P}}\geq \frac{10^{-30}}{\lambda ^{2}[\mathrm{m}]}%
\left( 1-\mathrm{e}^{-0.1/\lambda \lbrack \mathrm{m}]}\right) ^{-1}\left( 1-%
\mathrm{e}^{-d/\lambda }\right) ^{-1}.  \label{limit}
\end{equation}%
Figure 1 shows this sensitivity limit on $g_{\mathrm{S}}g_{\mathrm{P}}$,
together with limits provided by two other measurements and the proposed
limit of an upgraded gravitational level experiment. In order to establish
such a limit or even find a signal\footnote{%
A first experiment using the old apparatus of the Rutherford/Sussex/ILL
neutron EDM collaboration \cite{Baker/2006} is under way \cite{Pignol/2008}.}%
, a careful consideration of possible systematics is required. Here only
some general ideas are presented.

It is well known that instability and inhomogeneity of the magnetic field
may provide generous sources of systematic effects in neutron EDM searches
which partly are also relevant here. Past experiments have converged to the
strategy to let trapped neutrons precess in a $\mathrm{\mu T}$ field of a
solenoid, protected from external influences by several layers of magnetic
screen. Field monitoring has been performed with magnetometers using Cs in
proximity to the neutron Ramsey chamber \cite{Altarev/1992}, or using $%
^{199} $Hg as a co-magnetometer \cite{Baker/2006}. For the search of a
spin-dependent short-range force a neutron Ramsey chamber has to be
surrounded by magnetometers out of reach of the force, and $\omega _{\pm }$
in eq.(\ref{effect}) be measured as relative precession frequencies. Control
of global field drifts on a level required for a sensitivity on $g_{\mathrm{S%
}}g_{\mathrm{P}}$ as given in eq.(\ref{limit}) is met by state-of-the-art
magnetometry in EDM experiments, in particular using a double Ramsey chamber
setup mentioned further below. Note also that the problem of geometric
phases due to neutron motion in an applied electric field in conjunction
with magnetic field gradients \cite{Pendlebury/2004} is absent.%
\begin{figure}[ptb]\begin{center}
\includegraphics[
natheight=3.4379in, natwidth=4.3684in, height=3.61in, width=5.0718in]
{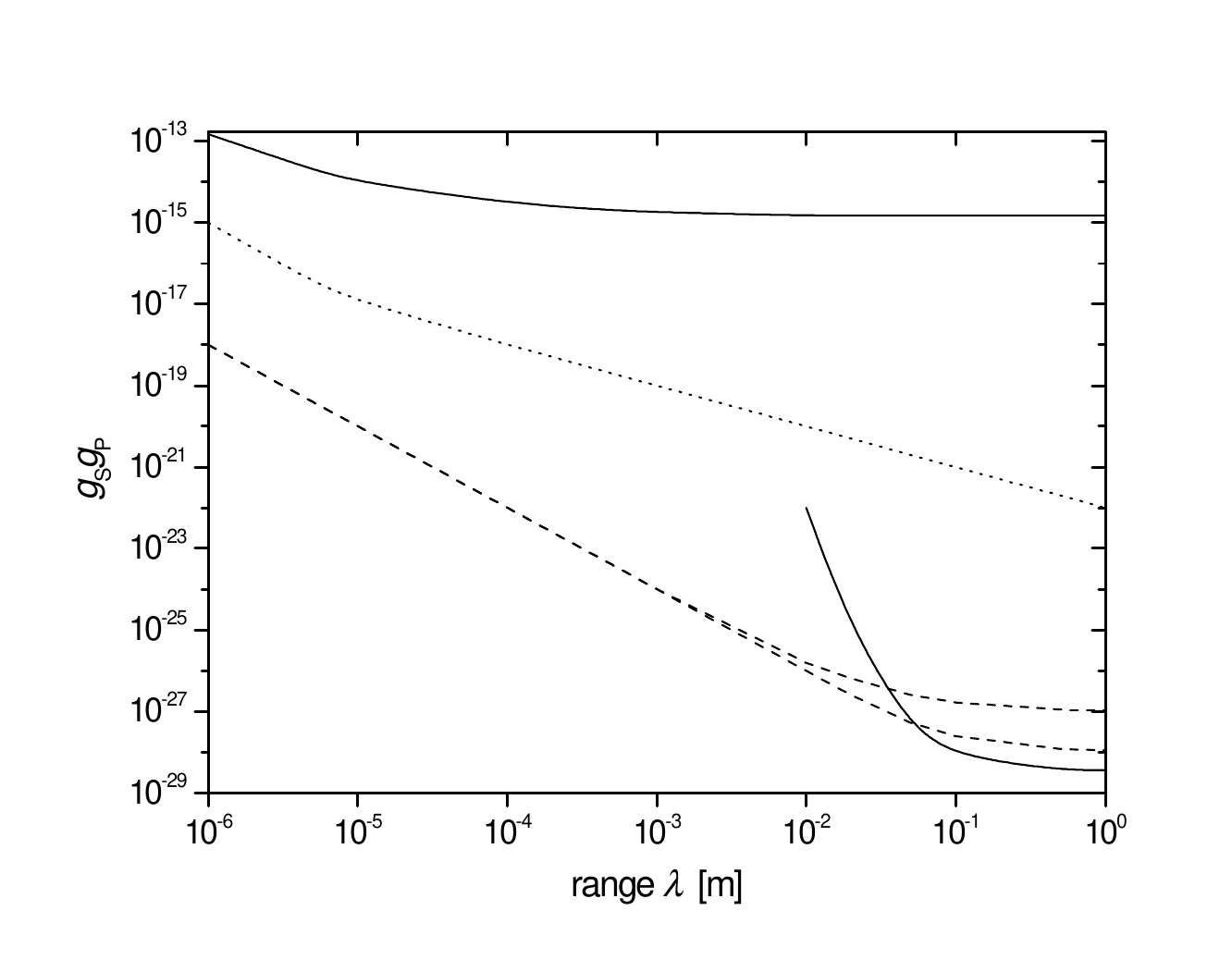}%
\caption{Constraints for the coupling constant product $g_{\mathrm{S}}g_{%
\mathrm{P}}$, as a function of range of the macroscopic force. The dashed
lines represents the proposed sensitivity limit of a spin precession
experiment (eq.(\protect\ref{limit}) with $d=0.01$ m (upper line), and $d=0.1
$ m (lower)). The dotted line is the limit proposed for an upgraded
gravitational level experiment \protect\cite{Baessler/2007}. The solid lines
are due to measurements (upper \protect\cite{Baessler/2007}, lower 
\protect\cite{Youdin/1996}).}
\end{center}\end{figure}%

In order to cover the axion window and probe forces even down to the $%
\mathrm{\mu m}$ range the test mass should be in close contact with the
precessing fermions. The experiment \cite{Youdin/1996} obtained an excellent
limit on $g_{\mathrm{S}}g_{\mathrm{P}}$ for $\lambda \gtrsim $ a few cm. It
was obtained by comparing relative precession frequencies of Hg and Cs
magnetometers as a function of the position of two $475$ kg lead masses with
respect to an applied magnetic field. A minimum distance $l$ between the
test mass and the magnetometer cells gives rise to an additional factor $%
\exp \left( -l/\lambda \right) $ in eq.(\ref{V-average}), which truncates
the range of best sensitivity to values $\lambda \gtrsim l$. For a gas of
ultracold neutrons close contact with the test mass can easily be
established by replacing one of the electrodes of an EDM chamber by a heavy
mirror.

Using the combination of a single neutron Ramsey chamber with atomic
magnetometers one may either invert the magnetic field or the trap
orientation to determine two relative neutron precession frequencies $\omega
_{\pm }$. Due to the presence of remanent magnetic materials for shielding
external fields, a field reversal may change field gradients which in turn
may affect the relative precession frequencies due to inappropriate spatial
averaging \cite{Kirch/2008}. Therefore it seems to be a better option to
keep the field constant and rotate the trap inside the magnetic shield. As
an alternative one may swap baths of mercury as variable test masses as
proposed in \cite{Youdin/1996}. In this variant the UCN trap should be made
of mechanically supported UCN-reflecting foils to allow for close contact of
UCN and mercury to avoid loss of sensitivity for small $\lambda $.

Best results might be obtained using a double-chamber with opposite mass
polarity with respect to the magnetic field. It offers a large degree of
intrinsic cancellation of magnetic field drifts. For the neutron EDM search
the concept of a double-chamber setup was developed and successfully
employed by the Gatchina group \cite{Altarev/1992}. To get good control over
field gradients, the neutron chambers should be sandwiched between two large
area magnetometers (or magnetometer arrays) \cite{Borisov/2000}. Higher
order magnetic field gradients may be corrected for using a stack of
mass-polar pairs of Ramsey chambers which could be supplemented by
mass-symmetric neutron chambers for magnetometry. A corresponding scheme was
proposed for the neutron\ EDM search by A. Serebrov (for latest update see 
\cite{Serebrov/2007}).

\end{document}